\documentclass[aps,prl,superscriptaddress,preprintnumbers,twocolumn,groupedaddress,showpacs]{revtex4}
\usepackage{graphicx}

\def\beq{\begin{equation}}
\def\eeq{\end{equation}}
\def\bea{\begin{eqnarray}}
\def\eea{\end{eqnarray}}
\newcommand{\cO}{{\mathcal O}}
\newcommand{\nn}{\nonumber}
\newcommand{\eerrr}{e^+e^- \to \gamma\gamma\gamma}

\newcommand{\GeV}{{\rm GeV}}

\newcommand{\MeV}{{\rm MeV}}
\begin{document}

\title{  Electroweak radiative corrections to triple photon production at the ILC  }
\author{ Zhang Yu$^a$, Li Wei-Hua$^b$, Duan Peng-Fei$^a$, Song Mao$^c$ and Li Gang $^c$\\
{\small  $^a$ City College, Kunming University of Science and Technology, }  \\
{\small  Kunming, Yunnan 650051, People's Republic of China}  \\
{\small  $^b$ Key Laboratory of Neutronics and Radiation Safety,\\
 Institute of Nuclear Energy Safety Technology, Chinese Academy of Sciences,}\\
{\small  Hefei, Anhui, 230031, People's Republic of China} \\
{\small $^c$ School of Physics and Material Science, Anhui University, \\
Hefei, Anhui 230039, People's Republic of China}}
\begin{abstract}
In this paper, we present the precision predictions for three photon production in the standard model (SM) at the ILC including the full next-to-leading (NLO) electroweak (EW) corrections, high order initial state radiation (h.o.ISR) contributions and beamstrahlung effects. We present the LO and the NLO EW+h.o.ISR+beamstrahlung corrected total cross sections for various colliding energy when $\sqrt s \ge 200\GeV$ and the kinematic distributions of final photons with $\sqrt s = 500\GeV$ at ILC, and find that the NLO EW corrections, the h.o.ISR contributions and the beamstrahlung effects are important in exploring the process $\eerrr$.
\end{abstract}


\maketitle

\par
\section{1. Introduction}
\label{Sec:intro}
After the infusive discovery of the Higgs boson by the ATLAS and CMS collaborations at the CERN Large Hadron Collider (LHC) \cite{Aad:2012tfa,Chatrchyan:2012xdj}, the next important goals of further collider experiments are to determine the nature of the discovered Higgs boson, investigate the predictions of the Standard Model (SM), and search the hints for physics in the beyond Standard Model (BSM). The multiple gauge boson productions are extremely essential in probing the self-coupling properties of the gauge boson. Especially, neutral gauge boson couplings provide a clean window to study new physics in the BSM, since the trilinear neutral gauge boson couplings (TNGCs), which test the gauge structure of the SM, and the quartic neutral gauge boson couplings (QNGCs), which may provide a connection to the mechanism of electroweak symmetry breaking (EWSB), vanish in the SM at tree level and any deviation from the SM prediction might be connected to the residual effect of EWSB. To directly measure the quartic gauge boson couplings, the investigations of triple gauge boson production are required. Precise predictions for such SM processes are important as the quantum corrections are often comparable to the BSM effects. Presently, the phenomenological results for all the triple gauge boson productions at hadron collider in the SM are available at next-to-leading order (NLO) in QCD  \cite{Lazopoulos:2007ix,Hankele:2007sb,Campanario:2008yg,Binoth:2008kt,Bozzi:2009ig,Bozzi:2010sj,
Baur:2010zf,Bozzi:2011wwa,Bozzi:2011en} and for $WWZ$ and $WZZ$ productions in the SM can be obtained at electroweak (EW) NLO \cite{Nhung:2013jta,Yong-Bai:2015xna}. At the $e^+e^-$ colliders, calculations at EW NLO have been  performed to $WWZ$, $ZZZ$, $Z\gamma\gamma$, $WW\gamma$ and $ZZ\gamma$ productions \cite{JiJuan:2008nn,Wei:2009hq,Boudjema:2009pw,Chong:2014rea,Yu:2013dxa,Yu:2015kbg}.

\par
In this paper, we present the full NLO EW corrections to the triple photon production at the $e^+e^-$ colliders, as well as the high order ISR contributions at the leading-logarithmic approximation in the structure function method. The reaction $\eerrr$ provides a background to Higgs boson production in association with a photon and has been measured with L3 detector at LEP to get limits on the anomalous $H\gamma\gamma$ and $HZ\gamma$ couplings \cite{Achard:2004kn}. It can also be used to present the limits on the QNGC $Z\gamma\gamma\gamma$ from the $e^+e^-$ colliders \cite{Gutierrez-Rodriguez:2013eya}. We quantify the improvement in the predictions at the total cross section with various collider energy and kinematic distributions to match the ILC experimental accuracy.

\par
The rest of this paper is organized as follows: in Sect. 2, we describe the details of the calculation, mainly the virtual and the real emission as well as the high order contribution. The numerical results are discussed in Sect. 3 and finally, we conclude in Sect. 4.

\par
\section{2. Calculation details}
In our calculation, we adopt the 't Hooft-Feynman gauge and apply {\sc FeynArts-3.7} package \cite{Hahn:2000kx} to automatically generate the Feynman diagrams. The corresponding amplitudes are subsequently reduced by using {\sc FormCalc-7.4} program \cite{Hahn:1998yk}. we neglect the contributions from the Feynman diagrams which involve the Higgs/Goldstone-electron-positron Yukawa couplings since the electron mass is tiny.

\par
The leading-order (LO) cross section for $\eerrr$ process is $\cO(\alpha^3)$. At NLO $\cO(\alpha^4)$, we encounter virtual as well as real emission contributions resulting from an additional photon.Virtual amplitudes are already at $\cO(\alpha^{5/2})$, hence only the interference of them with the LO Born amplitudes will contribute to the NLO level. The real emission process at NLO level comes from additional photon emissions from the LO processes. The ultra-violet (UV) divergences, coming from the virtual contributions, are regularized using the dimensional regularization scheme and can be removed through proper counter terms \cite{Ross:1973fp,Denner:1991kt}. The infra-red (IR) singularities coming from the real emission processes get cancelled with those coming from the virtual processes. We regulate the IR singularities by using infinitesimal fictitious photon mass and extract them from the emission corrections by employing the dipole subtraction (DS) method.

\par
In the calculation of one-loop Feynman amplitudes, we adopt the {\sc LoopTools-2.8} package \cite{Hahn:1998yk} for the numerical calculations of the scalar and tensor integrals, in which the $n$-point ($n\le 4$) tensor integrals are reduced to scalar integrals recursively by using Passarino-Veltman algorithm and the 5-point integrals are decomposed into 4-point integrals by using the method of Denner and Dittmaier \cite{Denner:2002ii}. In our previous work \cite{Chong:2014rea,Yu:2014cka,Yu:2015kbg}, we addressed the numerical instability originating from the small Gram determinant ($detG$) and scalar one-loop 4-point integrals \cite{Boudjema:2009pw}.
In order to solve these instability problems  in the numerical calculations, we developed the {\sc LoopTools-2.8} package, which can automatically switch to the quadruple precision codes in the region of small Gram determinants, and checked the results with ones by using {\sc OneLoop} package \cite{vanHameren:2009dr}to verify the correctness of our codes.

\par
In the DS method, an auxiliary function, which has the same singular structure pointwise in phase space with the squared amplitude of the real emission process, is subtracted to obtained IR finite results, which can be integrated numerically. In order to get final unchanged results, the subtracted term is added again after analytical integration over the bremsstrahlung photon space. The formalism of the dipole subtraction, which is a process independent approach, was first presented for QCD with massless unpolarized partons by Catani and Seymour \cite{Catani:1996jh,Catani:1996vz,Catani:2002hc} and subsequently was generalized to photon radiation off charged fermions with arbitrary mass by Dittmaier \cite{Dittmaier:1999mb}.
In our calculations, we directly use the general subtraction formalism in Ref.\cite{Dittmaier:1999mb}. To verify the correctness of our numerical calculation, we also check the independence on the cut parameter labelled $\alpha$ with $\alpha\in(0,1]$, which is introduced in Refs.\cite{Nagy:1998bb,Nagy:2003tz} to control the size of dipole phase space. We transfer the formulae for QCD with partons in the initial state in Ref.\cite{Nagy:2003tz} in a straightforward way to the case of photon emission off incoming leptons.

\par
Due to the smallness of the electron mass, the emission of photons collinear to the incoming electrons or positrons induces the quasi-collinear IR singularities, i.e., initial state radiation (ISR). The ISR quasi-collinear IR singularities can be partially cancelled by the virtual corrections. The left ones would lead to large radiative corrections of the form $\alpha^n\log^n(m_e^2/Q^2)$ at the leading-logarithmic (LL) level. To achieve an accuracy at the few $0.1\%$ level, the higher-order contributions from this part beyond NLO have to be taken into account. According to the mass-factorization theorem, the LL initial state QED corrections can be expressed as a convolution of the LO cross section with structure functions by using the structure function method \cite{Denner:2000bj,Beenakker:1996kt},
\bea
\label{eq:isr}
\int d\sigma_{\rm ISR-LL}&=&\int_0^1dx_1\int_0^1dx_2\Gamma_{\rm ee}^{\rm LL}(x_1,Q^2)\Gamma_{\rm ee}^{\rm LL}(x_2,Q^2)\nn\\
&\times& \int d\sigma(x_1p_{\rm e^+},x_2p_{\rm e^-}),
\eea
where $x_1$ and $x_2$ denote the fractions of the momentum carried by the incoming electron and positron just before the hard scattering, $Q^2$ is the typical scale where the hard scattering occurs chosen as the colliding energy $\sqrt s$ in our calculations
and the LL structure functions $\Gamma_{\rm ee}^{\rm LL}(x,Q^2)$ are detailed in Ref.\cite{Beenakker:1996kt} up to $\cO(\alpha^3)$. The LO and one-loop contributions must to be subtracted to avoid double counting when we add the Eq.(\ref{eq:isr}) to the NLO EW corrected result. The explicit expression for the subtracted terms are presented in Ref.\cite{Denner:2000bj}. In the following, the subtracted ISR effect is called the high order ISR (h.o.ISR) contribution beyond $\cO(\alpha)$, labelled as $\Delta \sigma_{\rm h.o.ISR}$.

\par
In order to achieve high luminosities at linear colliders, the bunches of electrons and positrons have to be very dense. Under these circumstances, the electrons undergo acceleration from strong electromagnetic forces from the positron bunch (and vice versa). Both particles may emit photons so that they lose energy and momentum before the interaction. This synchrotron radiation is called {\it beamstrahlung} and has a strong effect on the energy spectrum $D(x_1, x_2)$ of the colliding particles. The observable $e^+e^-$ cross sections will be changed as
\bea
\label{eq:bs}
\int d\sigma_{\rm BS}^{e^+e^-}&=&\int_0^1dx_1\int_0^1dx_2D_{e^+e^-}(x_1,x_2)\nn\\
&\times& \int d\sigma(x_1p_{\rm e^+},x_2p_{\rm e^-}),
\eea
The energy spectrum $D(x_1, x_2)$ depends strongly on the accelerator design and assumed beam parameters and can be obtained with {\sc Circe1} \cite{Ohl:1996fi} by using ILC accelerator design parameters in this paper. We define the beamstrahlung effects as $\Delta\sigma_{\rm BS} = \sigma_{\rm BS}-\sigma_{\rm LO}$.

In this paper, the total EW corrected results are defined as the summation of the LO cross section, NLO EW corrections, the h.o.ISR contributions and the beamstrahlung effects,
\beq
\sigma_{\rm EW}=\sigma_{\rm LO}+\Delta\sigma_{\rm NLO}+\Delta\sigma_{\rm h.o.ISR}+\Delta\sigma_{\rm BS}
\eeq
In order to analyze the origin of the NLO EW corrections clearly, we also calculate the NLO photonic (QED), originating from virtual photon exchange and real photon radiation, and purely weak relative corrections with $\Delta\sigma_{\rm NLO}=\Delta\sigma_{\rm NLO\,\, QED}+\Delta\sigma_{\rm weak}$. The corresponding relative corrections of various effects are defined as $\delta\equiv\frac{\Delta\sigma_x-\sigma_{{\rm LO}}}{\sigma_{{\rm LO}}} \,\,(x={\rm NLO, NLO\,\,QED, weak, h.o.ISR, BS})$

\par
\section{Numerical results}
\label{Sec:result}
\subsection{Input parameters and event selection criterion}
The relevant SM input parameters used in our calculation are taken as\cite{Agashe:2014kda}:
\begin{equation}\arraycolsep 2pt
\begin{array}[b]{lcllcllcl}
\alpha(0)&=&1/137.035999074,\\
 M_{W} & = & 80.385~\GeV,  \qquad M_{Z}  =  91.1876~\GeV, \\
 m_e &=& 0.510998928~\MeV, \quad m_{\mu}=105.6583715~\MeV, \\
 m_{\tau}&=&1.77682~\GeV,\\
 m_u &=& 66~\MeV, \quad \qquad m_d = 66~\MeV, \\
 m_c&=&1.2~\GeV,\qquad m_s=150~\MeV, \\
 m_t&=&173.21~\GeV, \qquad m_b=4.3~\GeV.
\end{array}
\label{SMpar}
\end{equation}
where the current masses of the light quark (all quarks except $t$-quark) can reproduce the hadronic contribution
to the photonic vacuum polarization\cite{Jegerlehner:2001ca}

\par
In our default setup, in order to exclude the inevitably infrared singularity at tree level, we require  kinematic cuts for final-state three photons
\begin{eqnarray}\label{cut}
E_{\gamma} \ge 8~\GeV,~~~|\cos{\theta_\gamma}|\le 0.995,
\end{eqnarray}
which are experimentally accessible \cite{Bartels:2012ex,Baer:2013cma}.
We order the final photons according to their transverse momenta. The hardest photon with maximum transverse momentum is denoted by $\gamma_1$. Like wise, $\gamma_2$ and $\gamma_3$ represent the second and third hardest photon, respectively.
If additional photon emission is present, any further phase-space cuts will only be applied to the three visible photons with highest $p_T$, while the forth is treated inclusively to ensure IR safety.

\par
\subsection{Total cross section}
\begin{figure}[htbp]
\includegraphics[angle=0,width=3.2in,height=2.4in]{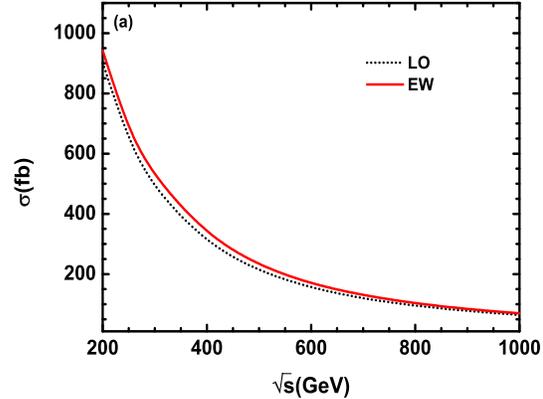}%
\hspace{0in}%
\includegraphics[angle=0,width=3.2in,height=2.4in]{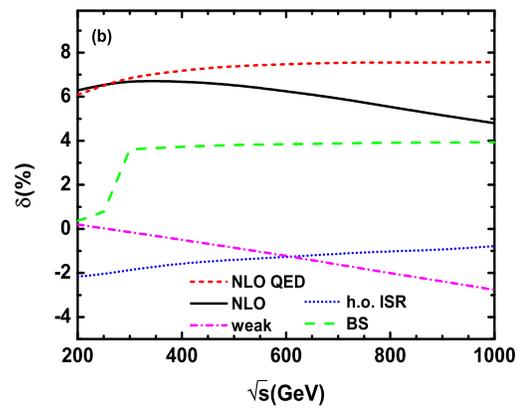}%
\hspace{0in}%
\caption{(a) The LO and total EW corrected cross sections ($\sigma_{{\rm LO}}$ and $\sigma_{{\rm EW}}$) for the $\eerrr$ process as the functions of the colliding energy $\sqrt{s}$ at the ILC. (b) The corresponding NLO EW, h.o.ISR, beamstrahlung, NLO QED and purely weak  relative corrections ($\delta_{{\rm NLO}}$, $\delta_{{\rm h.o.ISR}}$, $\delta_{\rm BS}$, $\delta_{\rm NLO\,\,QED}$ and $\delta_{\rm weak}$).} \label{fig-sqrts}
\end{figure}

In Fig.\ref{fig-sqrts}a, we present the LO and total EW corrected integrated cross section for the $\eerrr$ process with the colliding energy $\sqrt s \ge 200 \GeV$ in the SM at the ILC. The corresponding NLO EW, NLO QED, weak, h.o.ISR and beamstrahlung relative corrections are shown in Fig.\ref{fig-sqrts}b. Some representative numerical results of the LO and total EW corrected cross section, and the corresponding NLO EW, h.o.ISR and beamstrahlung relative corrections are presented in Tab.\ref{tab1}. From these figures, we find all the curves for the cross section decrease quickly with the increment of $\sqrt s$, and the LO cross sections are always enhanced by the NLO EW, beamstrahlung and total EW corrections  while reduced by the ISR effect beyond $\cO(\alpha)$ in the whole $\sqrt s$ range plotted. We also can see that the absolute NLO EW, h.o. ISR and beamstrahlung effects can maximally reach about $6.72\%$, $2.17\%$ and $3.93\%$ respectively, which are all notable. The weak relative corrections fall off with increasing $\sqrt s$, eventually reaching about $-2.8\%$ at an energy of 1 TeV. While the NLO QED relative corrections always rise and arrive at $7.6\%$ at 1 TeV. In the electroweak corrections, both QED and weak contributions partially compensate each other.
\begin{table}
\center
\begin{tabular}{cccccc}
\hline
$\sqrt{s}(\GeV)$ & $\sigma_{{\rm LO}}({\rm fb})$ & $\sigma_{{\rm EW}}({\rm fb})$
& $\delta_{{\rm NLO}}(\%)$& $\delta_{{\rm h.o.ISR}}(\%)$ & $\delta_{{\rm BS}}(\%)$\\
\hline \hline
200   & 902.08(47)   & 942.47(61)  & 6.29 & -2.17  &  0.36\\
250   & 639.84(33)   & 673.54(42)  & 6.54 & -2.04  &  0.77\\
300   & 480.03(25)   & 520.68(33)  & 6.72 & -1.86  &  3.61\\
400   & 301.70(16)   & 328.48(21)  & 6.70 & -1.56  &  3.73\\
500   & 208.92(11)   & 227.62(14)  & 6.54 & -1.40  &  3.80\\
600   & 154.10(8)    & 167.68(11)  & 6.25 & -1.28  &  3.84\\
800   & 94.72(5)     & 102.70(7)   & 5.54 & -1.02  &  3.91\\
1000  & 64.63(3)     & 69.77(5)    & 4.80 & -0.78  &  3.93\\
\hline  \hline
\end{tabular}
\caption{ \label{tab1}  The total LO cross section ($\sigma_{{\rm LO}}$), total EW corrected integrated cross sections ($\sigma_{{\rm EW}}$) and the corresponding NLO EW, h.o.ISR and beamstrahlung relative corrections ($\delta_{{\rm NLO}}$, $\delta_{{\rm h.o.ISR}}$ and $\delta_{\rm BS}$) for the $\eerrr$ process at the ILC.}
\end{table}

\par
\subsection{Kinematic distributions}
In this section we investigate some kinematic distributions of final photons for the $\eerrr$ reaction at the ILC with $\sqrt s=500\GeV$ including the NLO EW, h.o.ISR and beamstrahlung corrections.

\par
\begin{figure}[htbp]
\includegraphics[angle=0,width=3.2in,height=2.4in]{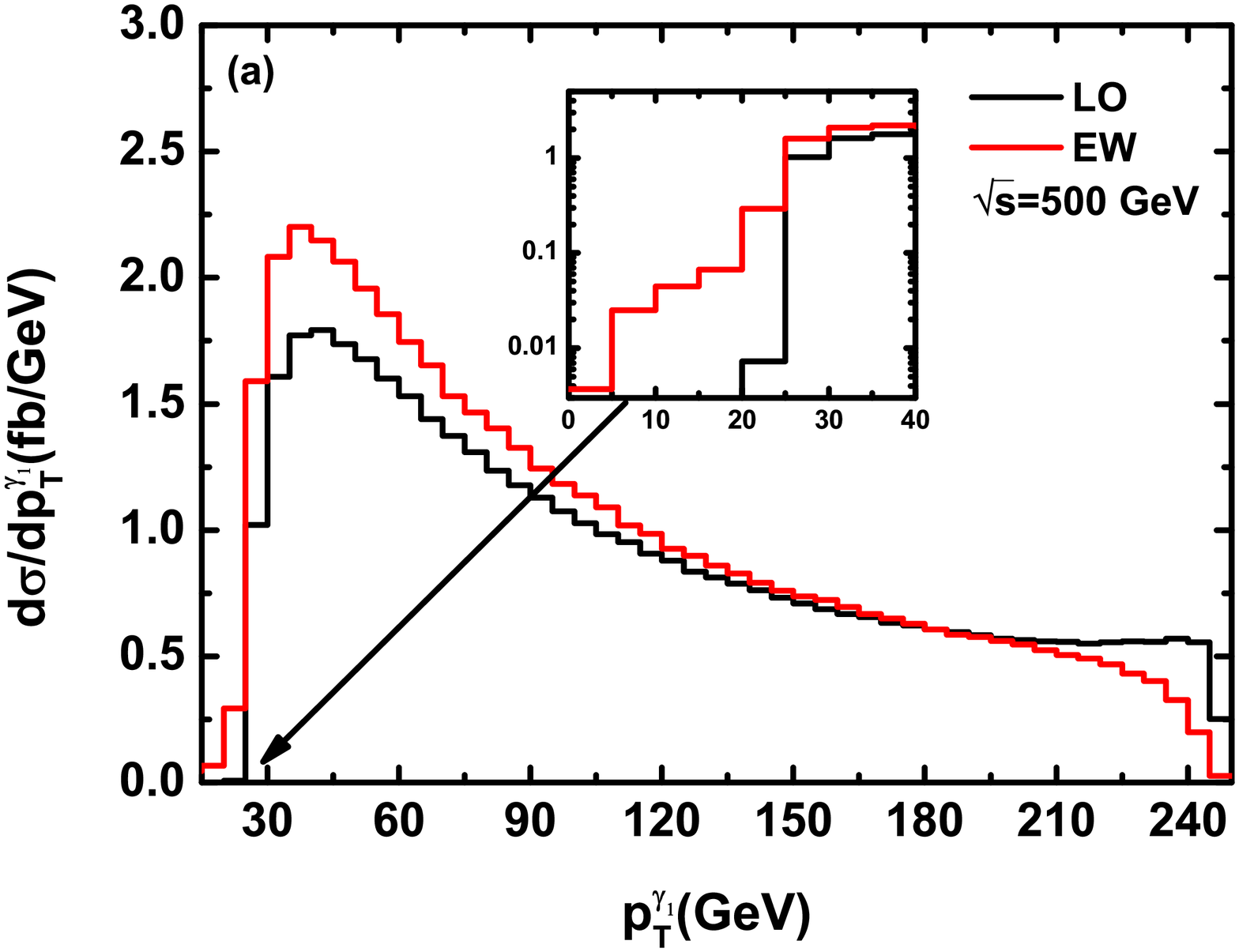}%
\hspace{0in}%
\includegraphics[angle=0,width=3.2in,height=2.4in]{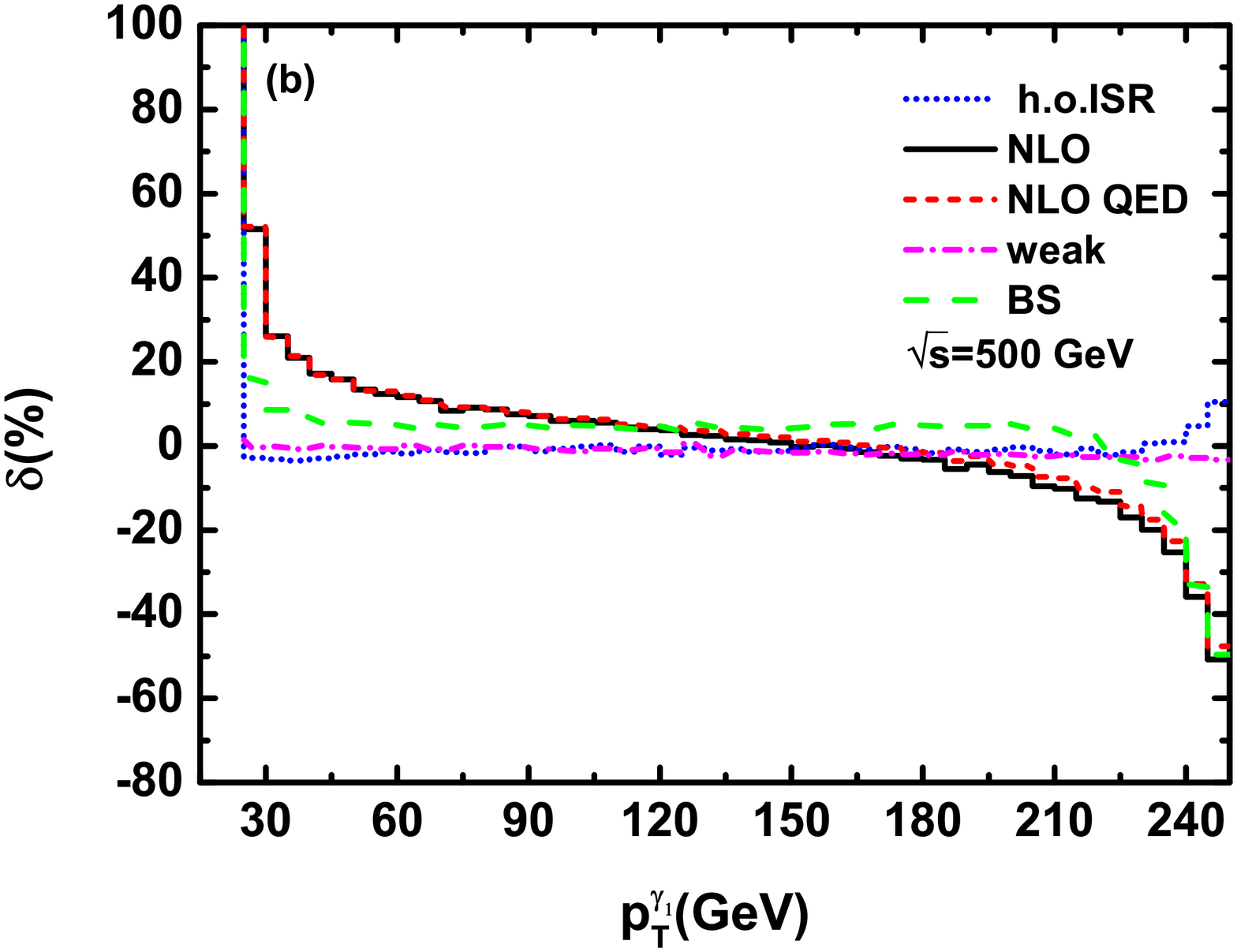}%
\hspace{0in}%
\caption{(a) The LO and total EW corrected transverse momentum distributions of the hardest photon $\gamma_1$ with $\sqrt s=500~\GeV$ at the ILC. (b)The corresponding NLO EW, h.o.ISR, beamstrahlung, NLO QED and purely weak relative corrections.}\label{fig-ptr}
\end{figure}

In Fig.\ref{fig-ptr}a, we show the LO and total EW corrected transverse momentum distributions of hardest photon $\gamma_1$ (i.e., $\frac{d\sigma_{LO}}{dp_T^{\gamma_1}}$ and$\frac{d\sigma_{EW}}{dp_T^{\gamma_1}}$). The bin-by-bin distributions of the NLO EW, h.o.ISR and beamstrahlung relative correction for the corresponding observable are provided in Fig.\ref{fig-ptr}b. We find that, the lower limit on the the hardest photon transverse momentum is $p_T^{\gamma_1}=20\GeV$ at LO because of the cut of $|\cos{\theta_\gamma}|\le 0.995$, whereas at NLO and/or higher order it can be very small due to the recoil against the extra photons helps to fulfill the transverse momentum cut, which was not possible at LO. For the same reason, the NLO EW, h.o.ISR and beamstrahlung relative corrections are very large at the lower plotted region. We can also see that the peak of the $p_T^{\gamma_1}$ distributions at LO is at $p_T^{\gamma_1}\sim 45\GeV$ while for EW corrected distributions, it is at $p_T^{\gamma_1}\sim 40\GeV$. That is, the hardest photon tends to be softer owing to the additional photon emission when the EW corrections are included. Besides, near the upper end of the $p_T^{\gamma_1}$ spectrum, the NLO EW, the h.o.ISR and the beamstrahlung corrections become more and more sizable, and can respectively reach about $-50.7\%$, $10.5\%$ and $-49.6\%$, which are all notable. At the peak of the LO $p_T^{\gamma_1}$ distributions, the NLO EW, the h.o.ISR and the beamstrahlung relative corrections are $17.2\%$, $-2.9\%$ and $5.4\%$ separately, which are also all sizable. Therefore, the NLO EW, the h.o.ISR and the beamstrahlung corrections are all worth being taken into account to measure the triple photon production at the ILC. In Fig.\ref{fig-ptr}b, we also plot the NLO QED and purely weak relative corrections. It can be seen that the NLO QED relative corrections become more and more negative with increasing $p_T^{\gamma_1}$ because of the reduced phase space after the extra photon radiation off the initial state, which prohibits a cancellation of large IR-sensitive virtual corrections by the real corrections. It is now clear that the large negative corrections for high photon transverse momentum result from NLO QED corrections. For the chosen $\sqrt s$, the purely weak corrections depend only weakly on $p_T^{\gamma_1}$ and are only at the per-cent level. The smallness of the purely weak corrections is mainly due to the minor corrections at $\sqrt s=500~\GeV$, which can be seen in Fig.\ref{fig-sqrts}b.

\par
\begin{figure}[htbp]
\includegraphics[angle=0,width=3.2in,height=2.4in]{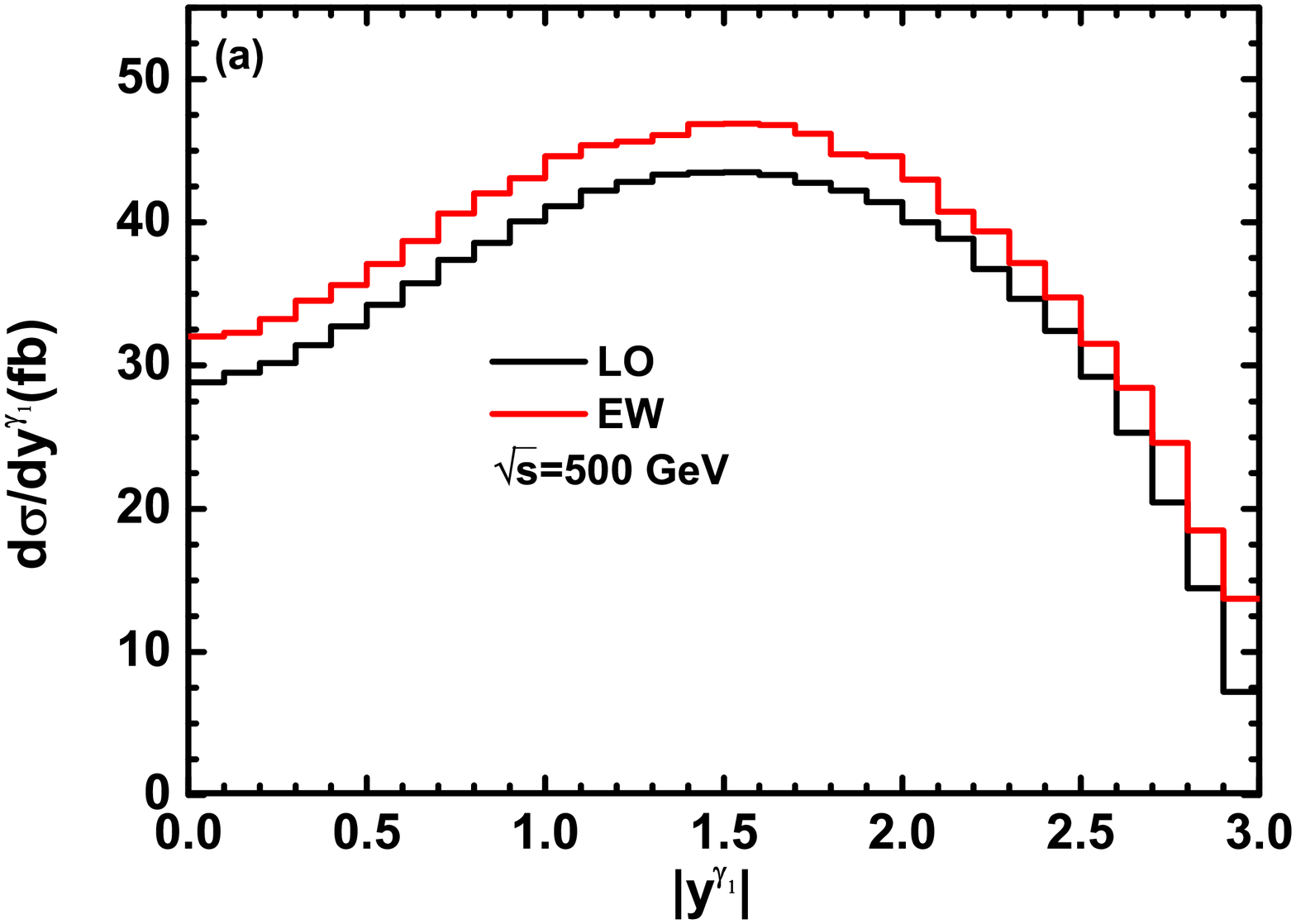}%
\hspace{0in}%
\includegraphics[angle=0,width=3.2in,height=2.4in]{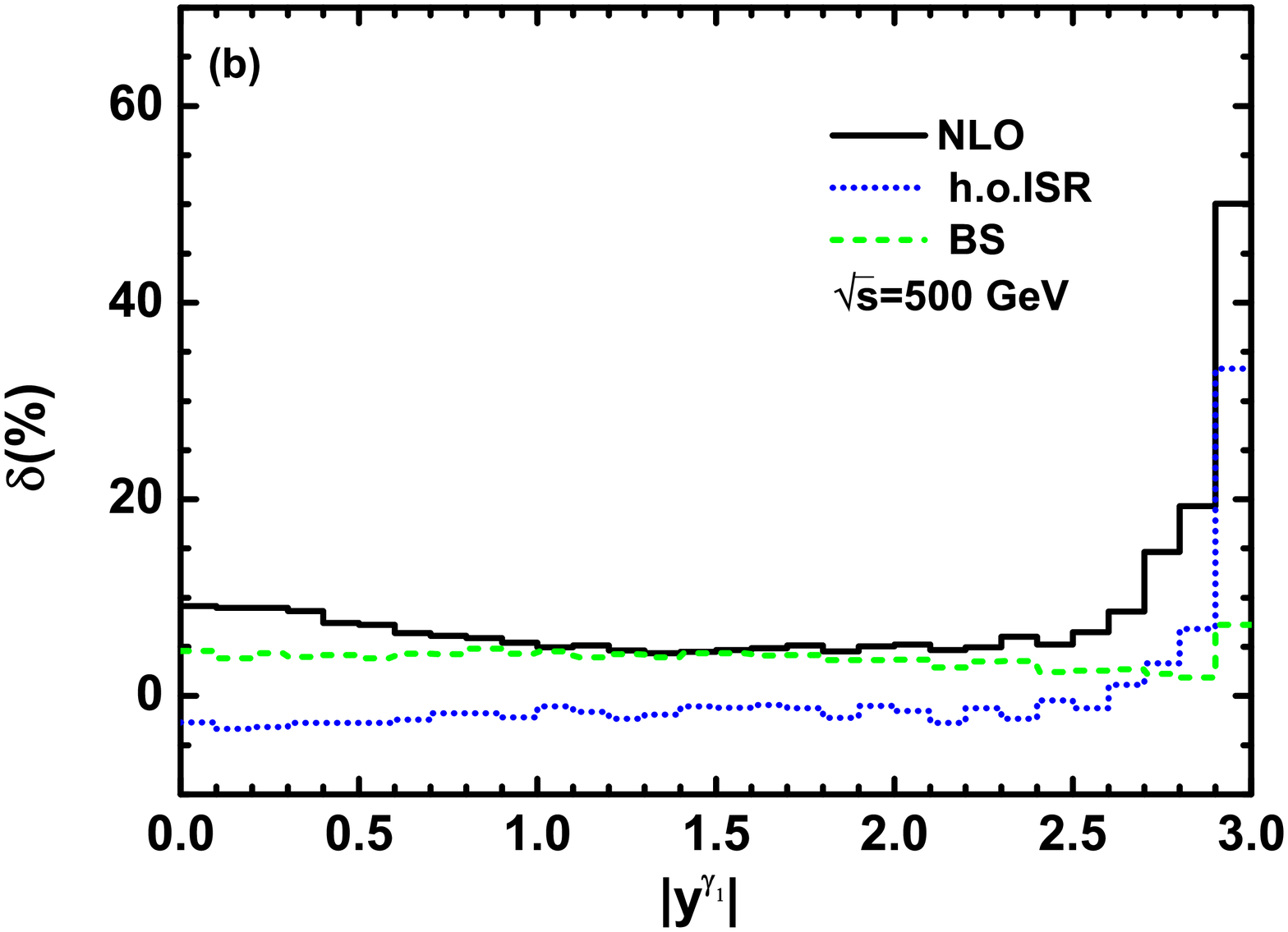}%
\hspace{0in}%
\caption{(a)The LO and total EW corrected rapidity distributions of the hardest photon $\gamma_1$ with $\sqrt s=500~\GeV$ at the ILC. (b) The corresponding NLO EW, h.o.ISR and beamstrahlung relative corrections.}\label{fig-rapr}
\end{figure}

The distribution of the rapidity of the hardest photon are shown in Fig.\ref{fig-rapr}. We can see that the LO and EW corrected rapidity distributions of the hardest photon peak at the position of $|y^{\gamma_1}|\sim1.6$ and the EW correction always increases the LO distribution in the whole plotted region. We see also that the absolute NLO EW, h.o.ISR, beamstrahlung relative corrections to rapidity distribution of the hardest photon reach their maximum at margin of the selection cut (i.e., $|y^{\gamma_1}|=3.0$).

\par
\begin{figure}[htbp]
\includegraphics[angle=0,width=3.2in,height=2.4in]{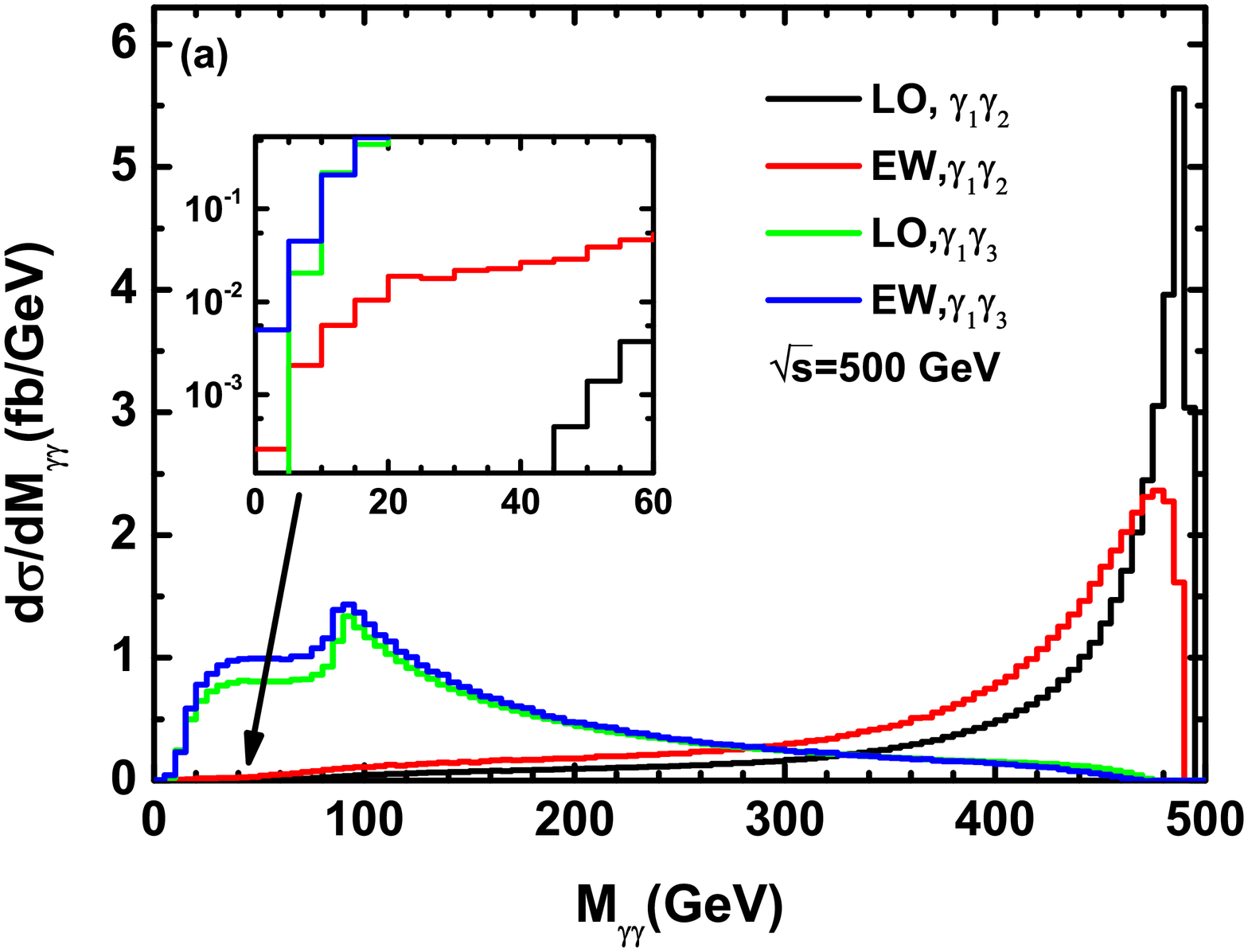}%
\hspace{0in}%
\includegraphics[angle=0,width=3.2in,height=2.4in]{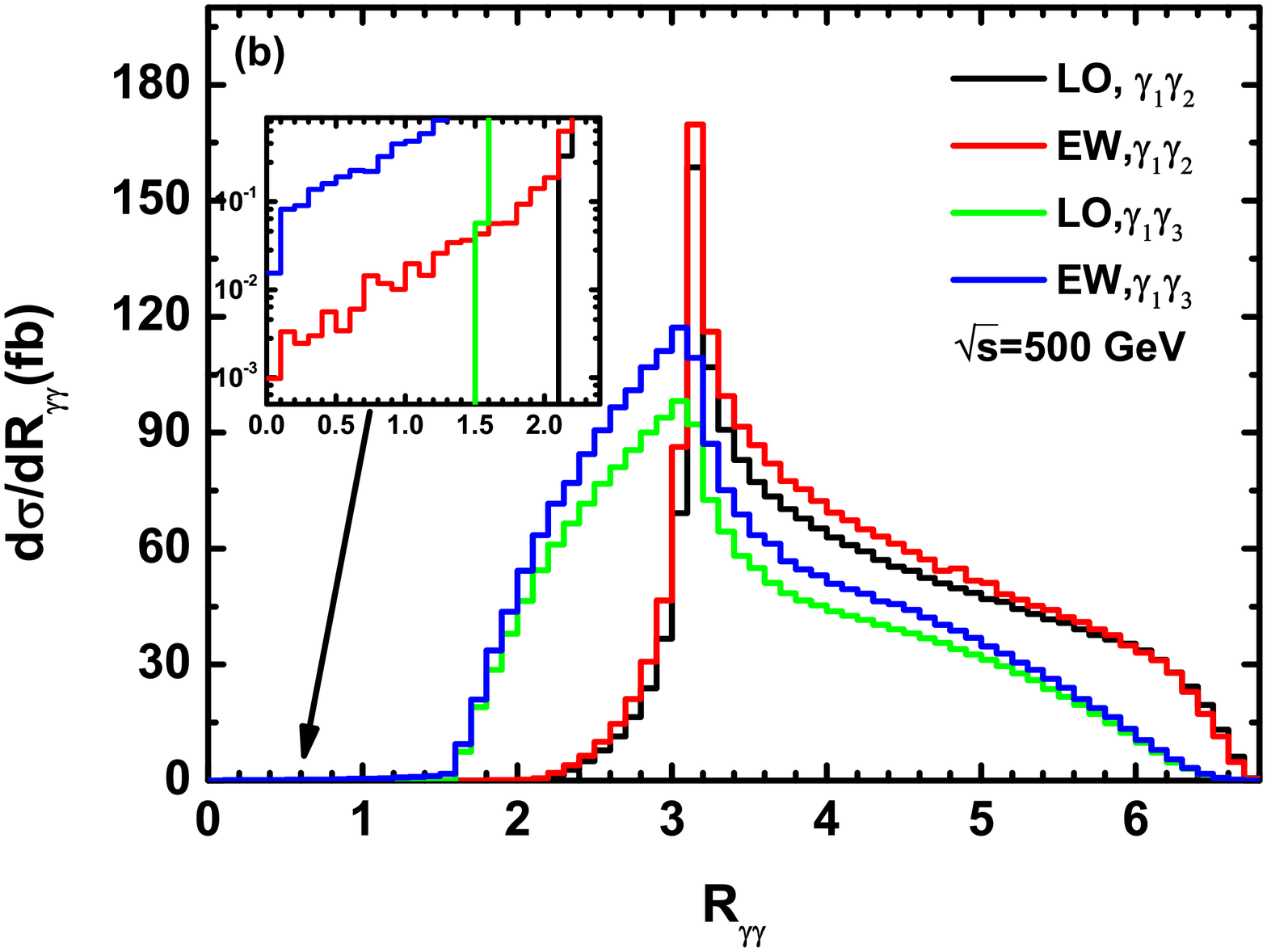}%
\hspace{0in}%
\caption{The LO and total EW corrected distributions for the invariant mass $M_{\gamma\gamma}$ (a) and the separation $R_{\gamma\gamma}$ (b) between the hardest photon $\gamma_1$ and the second/third hardest photon $\gamma_2/\gamma_3$ with $\sqrt s=500~\GeV$ at the ILC.}\label{fig-Mrr}
\end{figure}

The LO and EW corrected distributions of the invariant mass $M_{\gamma_i\gamma_j}$ and the separation in the rapidity and azimuthal angle ($y-\phi$) plane $R_{\gamma_i\gamma_j}$ ($R=\sqrt{\Delta y^2+\Delta \phi^2}$) between the ordered photons, where $i,j=1,2,3$, are plotted in Fig\ref{fig-Mrr}a and b respectively. The figures show that both the $M_{\gamma_1\gamma_2}$ and $M_{\gamma_1\gamma_3}$ tend to become smaller after total EW correct, which is natural since the additional EW radiation carries away energy. We also can see that the peaks arising in the $R_{\gamma_1\gamma_2}$ and $R_{\gamma_1\gamma_3}$ distributions near the angle $\pi$ ($180^\circ$), suggest that the emitted photons are mostly back-to-back. The hardest photon $\gamma_1$ is separated from the second (third) hardest one, i.e., $\gamma_2$ ($\gamma_3$), by at least $R_{\gamma_1\gamma_2}=2\pi/3$ ($R_{\gamma_1\gamma_3}=\pi/2$) at LO, whereas including the EW corrections they can be very small due to the emission of the extra radiations.

\par
\section{Summary}
In this paper, we present the full NLO EW corrections, the h.o.ISR contributions in the leading-logarithmic approximation and beamstrahlung effects to the triple photon production in $e^+e^-$ collision mode at the ILC. The $\eerrr$ process is very important background to understand the nature of the Higgs boson and explore the QNGC $Z\gamma\gamma\gamma$. We analyze the EW quantum effects on the total cross section and find that the LO cross sections are increased by the NLO EW, beamstrahlung and total EW  while reduced by h.o.ISR in the whole plotted $\sqrt s$ range, and the influences of the NLO EW, h.o.ISR and beamstrahlung on the total cross section are all sizable. We also investigate some important LO and EW corrected kinematic distributions of final photons, i.e., $p_T^{\gamma_1}$,  $y^{\gamma_1}$, $M_{\gamma_i\gamma_j}$ and $R_{\gamma_i\gamma_j}$, and find that the EW correction exhibits a strong dependence on the observable and on phase space. We conclude that both the NLO EW, h.o.ISR and beamstrahlung corrections are worth being taken into account in precision measurement of the triple photon production at the ILC.

\par
\section{Acknowledgments}
We gratefully thanks to Chen Chong for many helpful discussions. This work was supported in part by the Startup Foundation for Doctors of Kunming University of Science and Technology
(Grant No.KKSY201556046, No.KKSY201356060), the National Natural Science Foundation of China (Grant No.11405076, No.11347101,No.11205003£¬No.11305001), and the Science Foundation of Yunnan Provincial Education Department (Grant No.2014Y066).

\end{document}